\documentclass[aps,prb, groupedaddress,twocolumn,longbibliography]{revtex4-1}

\usepackage{graphicx, subfigure}
\usepackage{amsmath}
\usepackage{amsfonts}
\usepackage{amssymb}
\usepackage{bbm}

\usepackage{subfigure}

\usepackage{color}

\begin{document}


\global\long\def\ket#1{|#1\rangle}
\global\long\def\bra#1{\langle#1|}

\global\long\def\bd{b^{\dagger}}
\global\long\def\unit{1\!\!1}

\global\long\def\dt{\Delta t}
\global\long\def\rb{\bar{\rho}}

\title{Path integral description of combined Hamiltonian and non-Hamiltonian dynamics in quantum 
dissipative systems}

\author{A.M.~Barth$^{1}$, A.~Vagov$^{1}$, and V.M.~Axt$^{1}$}
\affiliation{$^{1}$Institut f\"{u}r Theoretische Physik III, Universit\"{a}t Bayreuth,
95440 Bayreuth, Germany,}
\email[]{andreas.barth@uni-bayreuth.de}

\date{\today}

\begin{abstract}

We present a numerical path-integral iteration scheme for the low dimensional reduced density matrix
of a time-dependent quantum dissipative system. Our approach simultaneously accounts 
for the combined action of a microscopically modelled pure-dephasing type coupling to a continuum
of harmonic oscillators representing, e.g., phonons, and further environmental interactions inducing
non-Hamiltonian dynamics in the inner system represented, e.g., by Lindblad type dissipation
or relaxation. Our formulation of the path-integral method
allows for a numerically exact treatment of the coupling to the oscillator modes and moreover is
general enough to provide a natural way to include Markovian processes that are sufficiently 
described by rate equations. We apply this new formalism to a model of a single semiconductor
quantum dot which includes the coupling to longitudinal acoustic phonons for two cases:
a) external laser excitation taking into account a phenomenological radiative decay of the excited 
dot state and
b) a coupling of the quantum dot to a single mode of an optical cavity taking into account cavity 
photon losses.

\end{abstract}

\maketitle


\section{Introduction}

Practically every quantum system experiences some kind of coupling to its environment and in many
cases a realistic modelling requires the inclusion of quantum dissipative 
processes\cite{weiss:99,breuer:02}. Such interactions with the environment typically lead to a
decay of quantum mechanical coherence within the subsystem of interest that is known as decoherence
or dephasing and often affects the dynamics in a non-negligible way. When the system-bath coupling
becomes strong or when the environmental correlation decay becomes slow it can be insufficient to
treat the environment as a constant entity simply acting on the system but instead the reaction of 
the external degrees of freedom to the system dynamics also has to be considered. Accounting for
these non-Markovian effects in a complete and correct way is not an easy task as besides the system
dynamics also the finite bath-memory has to be incorporated in the equations of motion. 
A powerful and widely used method that allows such an exact treatment is provided by the 
path-integral approach\cite{leggett:87, makri:95a, makri:95b, makri:14} which exactly takes into 
account the environment excitations via so called influence functionals for the degrees of freedom 
of the quantum system\cite{feynman:63}. 
This formalism has been applied in a variety of fields of both physics and chemistry such as energy 
transfer dynamics\cite{nalbach:11, nalbach:10b, nalbach:10c, eckel:09, thorwart:09, nalbach:14, 
kim:10, lee:12, liang:10}, Landau-Zener transitions\cite{nalbach:09, javanbakht:15}, quantum 
mechanical Brownian motion\cite{thorwart:00} and semiconductor quantum dots with and without 
optical driving\cite{thorwart:05, vagov:07, mccutcheon:11, vagov:11, glaessl:11b, glaessl:13b}. 
Moreover, it has been applied to systems with bosonic and fermionic baths\cite{segal:10, simine:13}, 
Ohmic and super-Ohmic\cite{vagov:11} environments and can also be used to include multiple baths.
However, in some cases the path-integral approach becomes impractical, because depending on the type
of environmental interaction the influence functional cannot always be obtained easily. On the 
other hand a completely microscopic description of the environment is not always necessary because
many dissipative processes are well known to be correctly described in the Markov limit and a 
simplified or even parametric treatment of the environment is sufficient. In these cases a realistic
modelling is usually achieved by simply adding phenomenological non-Hamiltonian contributions to the
equations of motion of the reduced system  as it is done, e.g., in the Lindblad  
formalism\cite{breuer:02}. 

In this paper we show that in a new generalized formulation the framework of the path-integral
method allows to treat such non-Hamiltonian dynamics on equal footing with the Hamiltonian part of
the equations of motion and can therefore also be used for models that describe some parts of the
environment by phenomenological rates while fully accounting for all non-Markovian effects induced 
by other couplings. This is not obvious, because the path-integral method usually relies on
describing the dynamics in terms of the time-evolution operator which yields purely Hamiltonian
dynamics.
More specifically here we present the path-integral formalism for a finite dimensional 
system that exhibits arbitrary non-Hamiltonian relaxation and a coupling to an arbitrary number of
microscopic harmonic oscillator modes that is of the pure-dephasing type, i.e., the coupling does 
not induce transitions between the finite basis states. 
After defining the model and establishing our new formalism in Section~\ref{model} we apply 
it to a strongly confined semiconductor quantum dot coupled to a continuum of longitudinal acoustic 
phonons in Section~\ref{application}. 
The new method allows for an unbiased study of the interplay of the carrier-phonon coupling with the
presence of radiative decay that is due to external field modes even 
in the regime of high temperatures and strong driving that is 
presented in Section~\ref{application}-A. By exemplarily comparing the path-integral calculations 
with a Markovian Master equation we also show that the new formalism can serve as an important 
benchmark tool. In Section~\ref{application}-B we then study the dynamics of a quantum dot coupled 
to a single photon mode inside a micro-cavity and compare the results with a previously developed 
hybrid approach. Finally, Section~\ref{conclusions} summarizes and concludes the paper.


\section{Model and Numerical Method}\label{model}  

Our generic model consists of a few level system with a pure-dephasing type coupling to a continuum
of harmonic oscillators. Further interactions between the few level system and the environment 
can be accounted for by additional non-Hamiltonian contributions to the equations of 
motion. The dynamics of the statistical operator for the total system consisting of the $N$ states 
of the few level system belonging to the subspace $\mathcal{H}_{N}$ and the oscillatory modes 
belonging to the subspace $\mathcal{H}_{osc}$ is given by the dynamical equation
\begin{equation}
\frac{d}{dt}\hat{\rho}=\frac{1}{i\hbar}\{\hat{H}, 
\hat{\rho}\}_{-}+\mathcal{L}[\hat{\rho}]\label{eq:masterequation}
\end{equation}
where $\{.,.\}_{-}$ denotes the commutator. The Hamiltonian 
\begin{equation}
\hat{H}=\hat{H}_{N}+\hat{H}_{osc} 
\end{equation}
consists of an arbitrary $N$-dimensional time-dependent Hamiltonian $\hat{H}_{N}$
acting only on the $N$ states $\{\ket{\nu}\}\in\mathcal{H}_{N}$ and
\begin{equation}
\hat{H}_{osc}=\hbar\sum_{j}\omega_{j}\hat{b}_{j}^{\dagger}\hat{b}_{j}+\hbar\sum_{\nu 
j}(\gamma_{j}^{\nu}\hat{b}_{j}^{\dagger}+\gamma_{j}^{\nu*}\hat{b}_{j})\ket{\nu}\bra{\nu}
\label{eq:hamiltonian_osc}
\end{equation}
which describes the interaction with the harmonic oscillators. Here, the symbol $\gamma_{j}^{\nu}$ 
denotes the coupling constant for the coupling between the state $\ket{\nu}$ and the bosonic mode
$j$ with energy $\hbar\omega_{j}$ that is created (destroyed) by $\hat{b}_{j}^{\dagger}$ 
($\hat{b}_{j}$). Note that for simplicity we treat the mode index $j$ as being discrete even though 
the presented formalism can account for an arbitrary number of modes. The pure-dephasing type of the 
coupling often dominates for systems where the states of the few level system are energetically well
separated such that inter-state transitions induced by the bosonic modes can be neglected.

While we denote Hilbert space operators with ``hat'' signs, the operator $\mathcal{L}[\hat{\rho}]$ 
appearing in Eq.~(\ref{eq:masterequation}) represents a Liouville space operator\cite{mukamel:95}, 
i.e. a linear mapping between such operators, that allows the inclusion of non-Hamiltonian 
dynamics. In the following we indicate such Liouville operators by putting the Hilbert space 
operator they act on into square brackets for clarity.
Here, $\mathcal{L}[\hat{\rho}]$ is assumed to keep the parts of $\hat{\rho}$ belonging to the 
subspace $\mathcal{H}_{osc}$ invariant and to only act on the few level system $\mathcal{H}_N$. We 
further assume that 
while $\mathcal{L}[\cdot]$ can be time-dependent, for simplicity it should be local in time. 
In principle, it would be possible to relax this assumption. However, depending on the memory depth 
of $\mathcal{L}[\cdot]$, this could increase the total memory time and thus the numerical cost of 
the path integral algorithm.
Moreover, in many cases it is advisable to require certain 
conditions for the operator $\mathcal{L}[\cdot]$ such that important physical properties of the 
density matrix are preserved. All of these conditions mentioned above are enforced when 
$\mathcal{L}[\cdot]$ has the so called Lindblad form\cite{breuer:02}
\begin{equation}
\mathcal{L}[\hat{\rho}]=\sum_{i}\gamma_{i}(t)\left(\hat{A}_{i}\hat{\rho} 
\hat{A}_{i}^{\dagger}-\frac{1}{2}\{\hat{\rho},\hat{A}_{i}^{\dagger}\hat{A}_{i}\}_{+}\right).
\label{eq:lindblad}
\end{equation}
where the operators $\hat{A}_{i}$ represent operations within $\mathcal{H}_{N}$, the factors 
$\gamma_{i}(t)$ denote possibly time-dependent relaxation rates, and $ \{.,.\}_{+}$ is the
anti-commutator. However, we would like to stress that the presented formalism does not depend 
on $\mathcal{L}[\cdot]$ to have this specific form.

Initially, the system is assumed to be in a product state of the states $\{\ket{\nu}\}$ and a 
thermal distribution of the oscillator modes $\hat{\rho}_{th}$, i.e., 
\begin{equation}
\hat{\rho}(t=0)=\hat{\bar{\rho}}(t=0)\otimes\hat{\rho}_{th}.
\end{equation}
The reduced density matrix of the $N$ level subsystem
\begin{equation}
\hat{\bar{\rho}}(t)=\mbox{Tr}_{osc}\left(\hat{\rho}(t)\right)
\end{equation}
is obtained by tracing out all the degrees of freedom belonging to the subspace $\mathcal{H}_{osc}$.
In the following we will derive a discretized representation of $\hat{\bar{\rho}}(t)$ that is 
applicable for numerical calculation and does not require further approximations to the model given 
above.

\subsection{Derivation of the path-integral solution}

We start by separating the right hand side of the master equation Eq.~(\ref{eq:masterequation}) into
two Liouville operators $\mathcal{L}_{N}[\cdot]$ and $\mathcal{L}_{osc}[\cdot]$ that are given by
\begin{eqnarray}
\mathcal{L}_{N}[\hat\rho] &=&  \frac{1}{i\hbar}\{\hat{H}_{N}, \hat\rho\}_{-} + 
\mathcal{L}[\hat\rho] \\ 
\label{eq:Phi}
\mathcal{L}_{osc}[\hat\rho] &=& \frac{1}{i\hbar}\{\hat{H}_{osc}, \hat\rho\}_{-}
\end{eqnarray}
and write the master equation as 
\begin{equation}
\frac{d}{dt}{\hat\rho}=\mathcal{L}_{N}[\hat\rho]+\mathcal{L}_{osc}[\hat\rho].
\label{eq:master_equation_Phi}
\end{equation}
Keeping the accuracy linear in a small time step $\dt$ the propagation of the statistical
operator to a time $t+\dt$ can be performed by applying $\mathcal{L}_{N}[\cdot]$ and 
$\mathcal{L}_{osc}[\cdot]$ subsequently and we can write
\begin{equation}
\hat\rho(t+\dt)=\hat{U}\mathcal{M}_{t}[\hat\rho(t)]\hat{U}^{\dagger} + \mathcal{O}(\dt^{2}). \\
\label{eq:single_step}
\end{equation}
Here, we have used the fact that the dynamics described by $\mathcal{L}_{osc}[\cdot]$ is
purely Hamiltonian and therefore can be expressed in terms of the time evolution operator
\begin{equation}
\hat{U}=\exp\left(-\frac{i}{\hbar}\hat{H}_{osc}\dt\right).
\end{equation}
Further, we have introduced the time-ordered operator
\begin{equation}
\mathcal{M}_{t}[\cdot]=\mathcal{T}\exp\left(\int_{t}^{t+\dt}\mathcal{L}_{N}dt^{\prime}\right)[\cdot]
\end{equation}
which acts as a generalized time evolution operator that describes the evolution of the few 
level system including the non-Hamiltonian part of the dynamics in the absence of the oscillator 
coupling from time $t$ to $t+\dt$. Importantly, the representation of these operators in terms of 
the matrix exponentials fulfills the necessary conservation requirements at each time step, 
such as a unitary evolution for the Hamiltonian dynamics.

We can now use the relation Eq.~(\ref{eq:single_step}) recursively to find an expression for the 
statistical operator at time $t$ starting from the initial time $t_0$. By inserting several 
identity operators $\sum_{\nu} \ket{\nu} \bra{\nu}$ acting on $\mathcal{H}_{N}$ at different time 
steps $t_l = t_0 + l\dt$ where the summation $\sum_{\nu}$ runs over all states of the few level 
system we arrive at a discretized representation
\begin{equation}
\hat\rho_{\nu_{n}\mu_{n}}=\sum_{\substack{\nu_{0}...\nu_{n-1} \\ 
\mu_{0}...\mu_{n-1}}}\hat{U}_{\nu_{n}}...\hat{U}_{\nu_{1}}\hat\rho_{\nu_{0}\mu_{0}}\hat{U}^{\dagger}
_{\mu_{1} } ...\hat{U}^{\dagger}_{\mu_{n}} 
\prod_{l=1}^{n}\mathcal{M}_{\nu_{l}\mu_{l}}^{\nu_{l-1}\mu_{l-1}}
\label{eq:discretized_representation}
\end{equation}
for the matrix elements of the statistical operator which we denote by
\begin{equation}
\hat\rho_{\nu_{l}\mu_{l}}=\bra{\nu_{l}}\hat\rho(t_l)\ket{\mu_{l}}.
\end{equation}
We have written 
\begin{equation}
\hat{U}_{\nu}=\bra{\nu}\hat{U}\ket{\nu}
\end{equation}
for the diagonal elements of the time-evolution operator $\hat{U}$ that are operators acting on the 
subspace $\mathcal{H}_{osc}$ and also introduced the symbol
\begin{equation}
\mathcal{M}_{\nu_{l}\mu_{l}}^{\nu_{l-1}\mu_{l-1}}=\bra{\nu_{l}}\mathcal{M}_{t}\left[\ket{\nu_
{ l-1 } }\bra{\mu_{l-1}}\right]\ket{\mu_{l}}
\end{equation}
which represents the matrix elements of the operator that results when $\mathcal{M}_{t}$ is applied 
to the canonical basis operator $\ket{\nu_{l-1}}\bra{\mu_{l-1}}$. 

We can now use the results of previous work\cite{vagov:11} where using a path-integral method and a 
representation of the oscillatory modes in terms of coherent states the trace over the 
oscillators for an operator of the same form as the one in Eq.~(\ref{eq:discretized_representation})
has been performed. Tracing out in this way the oscillator degrees of freedom we finally obtain the 
elements of the reduced density matrix $\hat{\bar{\rho}}$ at the $n$-th time step
\begin{multline}
\bar{\rho}_{\nu_{n}\mu_{n}}=\mbox{Tr}_{osc}\big{(}\hat{\rho}_{\nu_{n}\mu_{n}}\big{)} = \\ 
\sum_{\substack{\nu_{0}...\nu_{n-1} \\ 
\mu_{0}...\mu_{n-1}}}\bar{\rho}_{\nu_{0}\mu_{0}}\exp(S_{\nu_{n}...\nu_{1}}^{\mu_{n}...\mu_{1}}
)\prod_ { l=1 }
^{n}\mathcal{M}_{\nu_{l}\mu_{l}}^{\nu_{l-1}\mu_{l-1}}\label{eq:final_expression}
\end{multline}
where all summations run over the $N$ states of $\mathcal{H_N}$ and the influence functional 
$S_{\nu_{n}...\nu_{1}}^{\mu_{n}...\mu_{1}}$ incorporates the memory of the oscillator modes in the 
dynamics of the few-level system. For simplicity, here, we only give the expression for the 
influence functional for the special case where the coupling constants $\gamma_{j}^{\nu}$ are all 
either purely real or purely imaginary. A more general representation for the influence functional 
that does not make use of this assumption can be found in Ref.~\onlinecite{vagov:11}.
The functional reads
\begin{equation}
S_{\nu_{n}...\nu_{1}}^{\mu_{n}...\mu_{1}}=\sum_{l=1}^{n}\sum_{l^{\prime}=1}^{l}S_{\nu_{l}\nu_{l^{\prime}}\mu_{l}\mu_{l^{\prime}}}
\end{equation}
with
\begin{multline}
S_{\nu_{l}\nu_{l^{\prime}}\mu_{l}\mu_{l^{\prime}}}=-K_{\nu_{l^{\prime}}\nu_{l}}(t_{l}-t_{l^{\prime}}
)-K_{\mu_{l}\mu_{l^{\prime}}}^{*}(t_{l}-t_{l^{\prime}}) \\ 
+ K_{\nu_{l}\mu_{l^{\prime}}}^{*}(t_{l}-t_{l^{\prime}})+K_{\nu_{l^{\prime}}\mu_{l}}(t_{l}-t_{l^{
\prime }})
\end{multline}
and the memory kernels 
\begin{multline}
K_{\nu_{l}\mu_{l^{\prime}}}(\tau) = 
2\int_{0}^{\infty}d\omega\frac{J_{\nu_{l}\mu_{l^{\prime}}}(\omega)}{\omega^{2}}
(1-\cos(\omega\dt)) \\ \times
\left(\coth(\frac{\hbar\omega}{2k_{B}T})\cos(\omega\tau)-i\sin(\omega\tau)\right)
\label{eq:memory_kernel1}
\end{multline}
and
\begin{multline}
K_{\nu_{l}\mu_{l}}(0) = \int_{0}^{\infty}d\omega\frac{J_{\nu_{l}\mu_{l}}(\omega)}{\omega^{2}} 
 \\ \times
\left(\coth(\frac{\hbar\omega}{2k_{B}T})(1-\cos(\omega\dt))+i\sin(\omega\dt)-i\omega\dt\right)
\label{eq:memory_kernel2}
\end{multline}
where we have introduced the spectral density

\begin{equation}
J_{\nu\mu}(\omega)=\sum_{j}\gamma_{j}^{\nu}\gamma_{j}^{\mu*}\delta(\omega-\omega_{j}).
\label{eq:spectral_density_general}
\end{equation}
It should be noted that the last term in Eq.~(\ref{eq:memory_kernel2}) induces a polaronic shift of
the energy levels of the few level system.

\subsection{Evaluation of the path-integral expression}

Obtaining the reduced density matrix at the $n$-th time step from Eq.~(\ref{eq:final_expression}) 
requires the summation over $N^{2n}$ single contributions, which quickly becomes unfeasible even for 
very small $N$. Each of the single summands represents a possible {\it path}, i.e., a trajectory 
through the subspace $\mathcal{H}_N\otimes\mathcal{H}_N$ given by the configurations at each 
time-step $(\nu_{n},\mu_{n})...(\nu_{1},\mu_{1})$, which is the reason that the summation scheme is 
called a numerical path-integration method. To efficiently use this expression also for the 
iteration over many time-steps it is necessary to exploit the finite memory time of the system of 
oscillator modes that is reflected by a finite decay time of the memory kernels 
[Eqs.~(\ref{eq:memory_kernel1}, \ref{eq:memory_kernel2})]. This allows a truncation of the 
influence functional $S_{\nu_{n}...\nu_{1}}^{\mu_{n}...\mu_{1}}$, efficiently making it only depend 
on the states of the $n_c$ most recent time-steps 
$(\nu_{n}, \mu_{n})...(\nu_{n-n_c+1}, \mu_{n-n_c+1})$. Such a truncation can be exploited by 
combining the path-integral method with the augmented density matrix approach\cite{makri:95a, 
makri:95b} which applies also in the present case when including non-Hamiltonian dynamics.
The augmented density matrix can be thought of as a $2n_c$ dimensional tensor of
weights for the different possible configurations of the most recent $n_c$ time steps and 
can be calculated iteratively in each time step following the relation
\begin{multline}
\rho_{\nu_n...\nu_{n-n_c+1}}^{\mu_n...\mu_{n-n_c+1}}=\mathcal{M}_{\nu_{n}\mu_{n}}^{\nu_{n-1}\mu_{n-1
}} \\ \times 
\sum_{\substack{\nu_{n-n_c}\\\mu_{n-n_c}}}\exp(S_{\nu_{n}...\nu_{n-n_c}}^{\mu_{n}
...\mu_{n-n_c}})\rho_{\nu_{n-1}...\nu_{n-n_c}}^{\mu_{n-1}...\mu_{n-n_c}}.
\label{eq:augmented_density_matrix}
\end{multline}
This explicit iteration yields a numerical effort that is linear in the total number of time-steps 
and requires the calculation and storage of only $N^{2n_c}$ (compared to $N^{2n}$) weights and thus 
removes the restriction to a limited number of iteration steps in Eq.~(\ref{eq:final_expression}) 
that we mentioned above.

The decay time of the memory kernels is determined by the spectral density of the harmonic 
oscillator coupling $J(\omega)$ which can be classified by its low frequency behavior into sub-Ohmic
coupling where $J(\omega)\sim\omega^a$ with $a<1$ as $\omega\rightarrow0$, Ohmic coupling where
$a=1$ and super-Ohmic coupling where $a>1$. The Ohmic case marks the borderline between a sub-Ohmic
environment inducing exponential relaxations and the super-Ohmic case which is characterized by
non-exponential typically only partial relaxations that entail a variety of non-Markovian dynamical
effects\cite{leggett:87}. The presented formalism can also deal with the latter 
super-Ohmic case\cite{vagov:11} that is realized in the examples in this article for the coupling 
of acoustic phonons in a crystal solid where $a=3$.

Finally, we would like to note that the inclusion of non-Hamiltonian dynamics in 
Eq.~(\ref{eq:masterequation}) does not increase the required memory depth $n_c$ compared to the 
case of purely Hamiltonian dynamics as long as the former part of the dynamics does not involve a 
memory time that is longer than the one induced by the harmonic oscillator coupling. Here, this is 
obviously fulfilled as the operator $\mathcal{L}[\cdot]$ in Eq.~(\ref{eq:masterequation}) is assumed 
to be local in time.


\section{Application: Dynamics of a semiconductor quantum dot}\label{application}  

In this section we utilize the extended path-integral formalism to calculate the 
dynamics of an optically coupled strongly confined semiconductor quantum dot (QD) and also compare
the results with those of some established methods. For circularly polarized light with a central 
frequency close to the excitonic resonance a QD can be described for many purposes in good
approximation as a two-level system consisting of the crystal ground-state $\ket{0}$ and an exciton
state $\ket{X}$ with energy $\hbar\omega_X$ given by the Hamiltonian
\begin{equation}
 \hat{H}_{QD}=\hbar\omega_X\ket{X}\bra{X}
\end{equation}
Apart from the optically induced coherent dynamics which we will describe specifically in the 
corresponding two subsections also the coupling to the lattice vibrations of the surrounding solid 
state material needs to be taken into account. For the strongly confined GaAs-based QD considered 
here the deformation potential coupling to longitudinal acoustic (LA) phonons represents the 
predominant impact of the phonon environment\cite{krummheuer:02} and has the same form as
Eq.~(\ref{eq:hamiltonian_osc}) where the index $j$ refers to the wave vector $\bf q$ of the LA
phonon modes. The coupling constants $\gamma_{\bf q}^{\nu}$ are given by 
$\gamma_{\bf q}^{0} = 0$ for the ground state and
$\gamma_{\bf q}^{1} = \gamma_{\bf q}^{e} - \gamma_{\bf q}^{h}$
for the exciton state with
\begin{equation}
  \gamma_{\bf q}^{e(h)} = \Psi^{e(h)}({\bf q})\frac{|{\bf q}|D_{e(h)}}{\sqrt{2V\rho\hbar\omega_{\bf q}}}  
\end{equation}
being the coupling constants for the electron (e) and hole (h) coupling to the $\bf q$ phonon mode.
Here, the formfactors $\Psi^{e(h)}$ are assumed to be spherically symmetric and Gaussian as applies 
for a parabolic confinement potential and the deformation potential constants $D_{e}=7.0$~eV and 
$D_{h}=-3.5$~eV, the mass density $\rho=5370$~kg/m$^3$ as well as the sound velocity $c_s=5110$~m/s 
for GaAs are taken from the literature\cite{krummheuer:05}. The mode volume $V$ simply 
represents a normalization constant for the summation over the phonon modes in Eq. 
(\ref{eq:hamiltonian_osc}).
The spectral density of the phonon coupling given in Eq.~(\ref{eq:spectral_density_general}) is only 
non-vanishing for $\nu=\mu=1$ and in this case reads
\begin{equation}
 J_{11}(\omega)=\frac{\omega^3}{4 \pi^2 \rho \hbar c_{s}^5}\left(D_e e^{-\omega^2 a_e^2 / 
(4c_s^2)} - D_h e^{-\omega^2 a_h^2 / (4 c_s^2)} \right)^2
\end{equation}
where the sound velocity enters via the linear phonon dispersion relation 
$\omega_{\bf q}=c_s |{\bf q}|$ and $a_{e(h)}$ denote the root mean square of the Gaussian wave 
function extensions of the electron and hole, respectively. Here we set $a_e=4.0$~nm which can be
interpreted as the QD radius and set $a_e/a_h$ to 1.15. 

We would like to point out that there have been many suggestions to simulate the QD dynamics under 
the influence of the carrier-phonon interaction outlined above including correlation 
expansions\cite{foerstner:03,kruegel:06}, analytical solutions for delta 
excitation\cite{vagov:02}, an exact diagonalization approach\cite{kaer:13}, quantum jump 
approaches\cite{renaud:14} and various forms of master equations\cite{nazir:08,ramsay:10,kaer:10,
debnath:12,eastham:13,ardelt:14,nazir:16} some of which account for contributions of arbitrarily 
high order in the dot-phonon coupling with the help of the polaron 
transformation\cite{mccutcheon:10,mccutcheon:11,roy:12,kaer:12,manson:16}. This variety of methods
is also a result of the many different optical excitation scenarios that are discussed for QDs which 
can range from weak cavity couplings to strong pulsed laser excitation. The path integral approach
presented here provides a numerical scheme that allows to deal with all of these situations without 
introducing further approximations to the model formulated above.

\subsection{Laser-driven quantum dot with radiative decay}\label{applicationA}

As a first example for an application of our new method we consider the QD dynamics that is driven
by an external laser field and affected by both the phonon-induced relaxation and the radiative 
decay of the exciton state that reduces the exciton lifetime. As the radiative decay is known to 
be reasonably described as a Markov process in a good approximation it can be included by a 
Lindblad contribution to the master equation. 
In the rotating frame the contribution to the Hamiltonian for the laser driven QD 
reads after applying the common dipole and rotating wave approximations
\begin{equation}
 \hat{H}_{\mbox{dot-light}} = \frac{1}{2}\hbar f(t)\big{(}\ket{0}\bra{X} + \ket{X}\bra{0}\big{)} - 
\Delta\ket{X}\bra{X}
\end{equation}
where $f(t)$ is the envelope function of the laser field referred to as field strength and
$\Delta$ is the detuning of the laser from the polaron-shifted exciton resonance.
The radiative decay of the exciton state is treated as a phenomenologically damping rate accounted
for by a Lindblad type operator in the form of Eq.~(\ref{eq:lindblad}) with a single transition
$\hat{A}_{1} = \ket{0}\bra{X}$ and a corresponding damping rate $\gamma$ (c.f. 
Ref.~\onlinecite{kaer:12}).
\begin{figure}[ttt]
 \includegraphics[width=8.7cm]{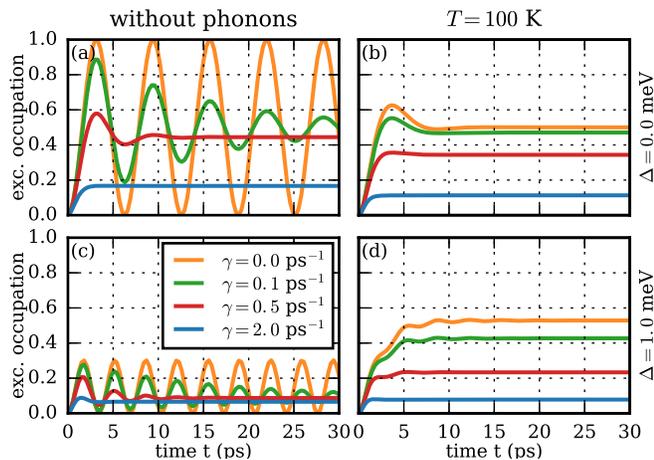}
    \caption{(Color online)
            Time dependent exciton occupation $C_X$ of a QD calculated by the path-integral formalism
            for constant driving with field strength $f=1.0$~ps$^{-1}$ for resonant [(a), (b)]
            and detuned [(c), (d)] excitation ($\Delta=1.0$~meV) for different values of the 
            radiative decay rate (see legend). Left column: without phonon
            interaction, right column: including phonon interaction at temperature $T=100$~K in
            addition to the radiative decay.
            }
 \label{fig1}
\end{figure}

Fig.~\ref{fig1} shows the time-dependent exciton occupation of the QD under resonant and 
off-resonant cw-excitation as indicated. In the absence of the phonon interaction (left column) our
computational scheme exactly reproduces known analytical results exhibiting
damped Rabi oscillations. For off-resonant excitation [c.f. panel (c)] of course the amplitude of 
the oscillations is reduced and the Rabi frequency is increased as can be seen in Fig.~\ref{fig1}(c).
The radiative decay also influences the stationary exciton occupation for $\gamma>0$, which is
given by\cite{loudon:73}
\begin{equation}
 C_{X,\rm{no~ph.}}^{\infty}=\frac{f^2}{2f^2+\gamma^2+(2\Delta/\hbar)^2}\label{eq:exciton_stationary}
\end{equation}
and decreases from its maximum value of $0.5$ with an increasing damping rate and an increasing
detuning. This simple application without the phonon interaction serves as a proof of principle that
the formalism presented here correctly incorporates non-Hamiltonian dynamics of Lindblad-type 
within the framework of the path-integral method.

In the other limiting case where $\gamma=0$, but the phonon coupling is included 
the results obtained from
the presented formalism coincide with previous path-integral calculations\cite{vagov:11}
that did not yet allow to account for any non-Hamiltonian dynamics. This can be seen explicitly
from the orange curve for $\gamma = 0$ in the right column of Fig.~\ref{fig1} where calculations 
that include the phonon coupling for a temperature of $T=100$~K are shown. As known from previous
simulations at such high temperatures the Rabi oscillations are almost completely damped by the
phonon scattering.

As for the radiative damping also the phonon coupling strongly affects the stationary 
value of the exciton occupation that is reached at long times. It can be seen that for fixed 
$\gamma$ the stationary value $C_{X}^{\infty}$ is slightly decreased by the phonon coupling for
resonant excitation while it is increased for off-resonant excitation. The reason for this is that
while the radiative decay always drives the QD towards the ground state and thus away from the
exciton, off-resonant excitation with positive detuning enables phonon-assisted transitions between 
the laser dressed 
states that yield a higher exciton occupation\cite{glaessl:11, hughes:13, glaessl:13, ardelt:14,
reiter:14, quilter:15, bounouar:15, manson:16}.
As it can be seen in Fig.~\ref{fig1} this
feature prevails when the phonon induced relaxation between photon dressed states is combined with 
the radiative decay of the exciton state discussed here.

\begin{figure}[ttt]
 \includegraphics[width=8.7cm]{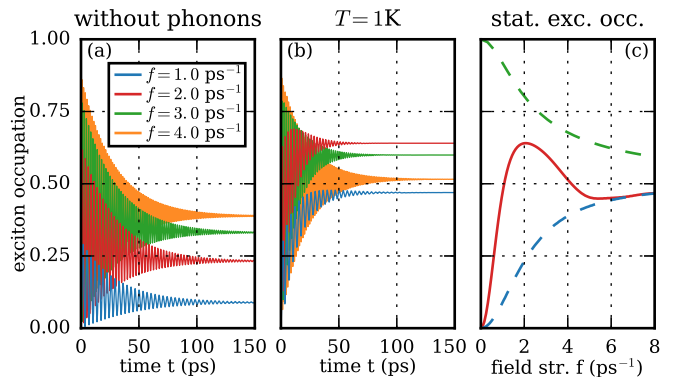}
    \caption{(Color online)
            Time-dependent exciton occupation $C_X$ of a QD for off-resonant ($\Delta=1.0$~meV)
            cw-excitation for different field strengths (see legend) including radiative decay
            ($\gamma=0.05~\mbox{ps}^{-1}$) without the phonon-interaction (a) and including phonons
            at temperature $T=1$~K (b). (c): Stationary exciton occupation reached at long times
            when only the phonon-interaction is present (green, dashed), when only the radiative
            decay is present (blue, dashed) and when both relaxation mechanisms are present
            (red, solid).
           }
 \label{fig2}
\end{figure}

Fig.~\ref{fig2} shows the combined influence of the radiative decay and the phonon-interaction by
plotting the time-dependent exciton occupation under positively detuned cw-excitation for different
field  strengths as indicated. Without the phonon interaction [panel (a)] the stationary exciton
occupation that is reachead at long times $C_{X}^{\infty}$ always stays below $0.5$ and rises with
increasing field strength as it was expected from Eq.~(\ref{eq:exciton_stationary}).
Interestingly, when including the phonon scattering [panel (b)] the stationary exciton occupation no
longer depends on the field strength in a monotonic way.
To analyze this in more detail we have plotted $C_{X}^{\infty}$ as a function of the field strength
in Fig.~\ref{fig2}(c) (red, solid) together with corresponding results for the two
limiting cases where only the phonon scattering (dashed, green) or only the radiative decay
(dashed, blue) has been accounted for. In case of the complete dynamics there is a clear maximum 
around $f = 2.0$~ps$^{-1}$ and a local minimum around $f = 5.0$~ps$^{-1}$ and overall the full model
predicts very different features compared to the two limiting cases. This behavior originates from 
the combination effects between the radiative decay and the phonon-induced relaxation as will be
explained in the following. For very small field strengths, e.g. below $f = 0.5$~ps$^{-1}$,
the timescale of the phonon-induced relaxation\cite{barth:16} is long
compared to the radiative decay rate and therefore one might expect the phonon coupling to play a
subordinate role. However, in this regime the QD state targeted by the phonon-induced 
relaxation has an especially strong excitonic character\cite{glaessl:11b}, which without 
radiative decay would result in an exciton occupation near one as can be seen by the dashed green 
line. When the radiative decay is included a remainder of this strong effect is still visible 
and thus leads to a clear difference between the results with (red, solid) and without 
phonons (blue, dashed) even for low field strengths. For larger values of $f$ the
phonon-induced relaxation becomes more effective yielding a steep increase of $C_X^{\infty}$ and 
values well above $0.5$ that would not be expected from the radiative decay alone.
Beyond the maximum at $f=2.0$~ps$^{-1}$ $C_X^{\infty}$ decreases again as the QD state targeted
by the phonon relaxation becomes less excitonic in character and even more importantly the 
phonon-coupling becomes less efficient again because the phonon environment is too sluggish to follow the
rapid dynamics of the QD\cite{vagov:07, ramsay:10b}. At very large $f > 5.0$~ps$^{-1}$ the phonon
coupling no longer has a significant impact on $C_X^{\infty}$ and the stationary exciton occupation
is almost entirely dominated by the radiative decay resulting in a local minimum and a subsequent 
slow increase towards $0.5$. It is worth noting that the non-monotonic dependence of the 
damping of the Rabi oscillations that is due to the phonon coupling persists under the influence of 
the radiative decay as it can be seen from Fig.~\ref{fig2}(b).
\begin{figure}[ttt]
 \includegraphics[width=8.7cm]{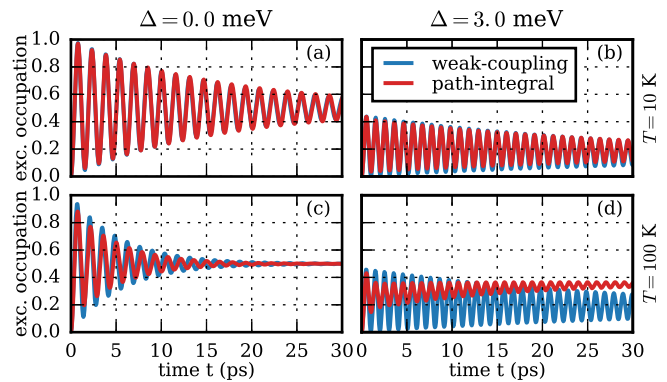}
    \caption{(Color online)
            Comparison of the time-dependent exciton occupation of a QD for resonant (left) and
            detuned (right) cw-excitation calculated by the path-integral method (red) and the weak
            coupling theory (blue). The radiative decay rate has been set to
            $\gamma = 0.05$~ps$^{-1}$ and a relatively high field strength of $f = 4.0$~ps$^{-1}$ 
            was chosen.
            }
 \label{fig3}
\end{figure}

A strong advantage of the numerical path-integral method is that because of the exact treatment of
the model discussed here it can be used to benchmark perturbative approaches and explore
their range of validity. To this end we have compared our results with those of a Markovian master
equation that treats the phonon coupling up to second order which we took from the
literature\cite{nazir:16} and that also allows an inclusion of the radiative exciton decay 
considered here. 
Fig.~\ref{fig3}(a) shows the damped Rabi oscillations of the exciton occupation under
constant resonant excitation by a strong electric field ($f=4.0$~ps$^{-1}$) under the influence 
of the phonon coupling ($T=10$~K) and the radiative decay ($\gamma=0.05$~ps$^{-1}$). As it can be
expected at such low temperatures the weak-coupling theory (blue) works very well and the predicted
dynamics of the exciton occupation is practically identical with those of the path-integral method
(red) which indicates that in this parameter range non-Markovian effects and multi-phonon processes
that the weak-coupling theory cannot capture are of minor importance.
Notably, the close match between the two methods is an important further verification of the
correctness of the presented path-integral algorithm with Lindblad-type relaxation.
When raising the temperature to $T=100$~K, c.f. Fig.~\ref{fig3}(c), some differences between the
results of the path-integral calculations and the weak-coupling theory become visible. 
The phonon-induced damping is slightly
underestimated by the weak-coupling theory and also there is a slight discrepancy of the predicted
Rabi frequency renormalization between the two methods. However, considering the very high 
temperature the Master equation still yields reasonable results. This is due to the strong driving
chosen here while for slower driving it is well known that the weak coupling theory can not account
for the strong Rabi frequency renormalization\cite{mccutcheon:11} and even can yield unphysical 
results at very high temperatures\cite{nazir:16}. For off-resonant excitation 
the path-integral calculations again practically coincide with those obtained from the Master 
equation approach at $T=10$~K [panel (b)]. However, at higher temperatures, c.f. 
Fig.~\ref{fig3}(d), strong differences between the two approaches become visible. The phonon-induced
damping is clearly underestimated by the weak-coupling theory and even more important the stationary
exciton occupation at long times predicted by the weak-coupling theory is considerably lower than
the value predicted by the path-integral calculations. The regime of strong driving at elevated 
temperatures is especially difficult to deal with in the Master equation approach. Because the
path-integral method does not rely on any approximations regarding the order of the phonon coupling
or the optical driving that it accounts for and is only limited by the errors introduced by the
discretization of the time axis it provides an important benchmark. Most importantly, the regime
of strong driving must be considered when simulating pulsed excitation scenarios in which high
field strengths can be reached and that are often required to reproduce experimental results.

\subsection{QD coupled to a single cavity mode}\label{ApplicationB}

Another system that can be described within the combined Lindblad and path-integral method is a QD
inside an optical cavity. Here, we assume that the quantized cavity photon modes are sufficiently
separated in frequency such that only a single mode effectively couples to the QD. Further, we 
assume that the system can be described in the single photon limit where only states with zero 
or one cavity photon have to be considered. Besides the coupling to LA phonons that is 
independent of the cavity coupling also photon
losses that are due to imperfections of the cavity mirrors are highly relevant for the system
dynamics. Similar to the previous examples we model the dynamics of the QD consisting of two
electronic levels coupled to the cavity mode and the phonon subsystem in an exact Hamiltonian way
while we attribute the cavity losses to the part of the environment that is described by rate 
equations. In the rotating frame the Hamiltonian for the QD coupled to the cavity mode is described 
via the Jaynes-Cummings model and reads after applying the common dipole and rotating wave
approximations
\begin{multline}
 \hat{H}_{\mbox{dot-cav}} = \hbar g \big{(}\ket{P}\bra{X}+\ket{X}\bra{P}\big{)} - 
\Delta\ket{X}\bra{X}
\end{multline}
where $g$ is the light-matter coupling strength, $\Delta$ is the detuning of the cavity-mode from the
polaron-shifted exciton resonance, and the two-level electronic basis of the QD-cavity system
consists of the state $\ket{P}$ with the QD in the ground-state and one cavity photon 
and the
exciton state $\ket{X}$ without a cavity photon. Only $\ket{X}$ couples to the 
phonon enivironment and the coupling is the same as in the previous examples.
The photon losses of the cavity are modelled by a relaxation with rate $\kappa$ from
$\ket{P}$ to the state where the QD is in its ground-state and no photon is present
$\ket{G}$ which is accounted for by a Lindblad contribution to the equation of motion
for the reduced density matrix\cite{kaer:12} by setting $\hat{A}_{1} = \ket{G}\bra{P}$ in 
Eq.~(\ref{eq:lindblad}). 

\begin{figure}[ttt]
 \includegraphics[width=8.7cm]{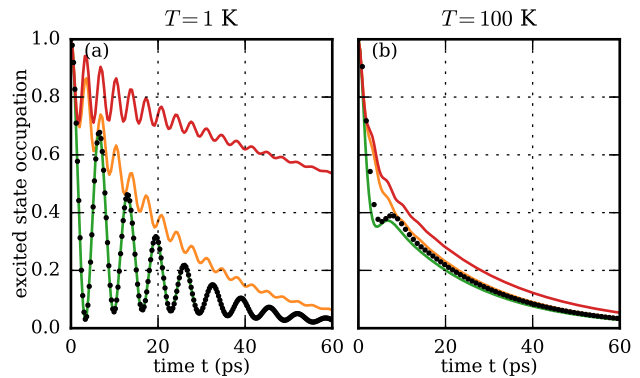}
    \caption{(Color online)
            Time-dependent occupation of the excited state $C_X$ of a QD inside a cavity for 
            resonant (green) and off-resonant (red: $\Delta=1.0$~meV, orange: $\Delta=-1.0$~meV) 
            coupling. The black dots show the results of a hybrid approach that is only applicable 
            in the resonant case (see text).
            Left panel: $T=1$~K, Right panel: $T=100$~K. $\kappa=0.1$~ps$^{-1}$.
            }
 \label{fig4}
\end{figure}

Fig.~\ref{fig4} shows the occupation of the excited state $\ket{X}$ as a function of time for a 
system initially prepared in the excited state for a cavity-loss rate $\kappa=0.1$~ps$^{-1}$ at two
different temperatures 
for resonant and off-resonant coupling (see caption). For resonant coupling we can see Rabi
oscillations with a decreasing amplitude, which can be attributed to the cavity losses. Similar 
oscillations can also be seen for off-resonant coupling, but in this case a positive detuning (red) 
leads to a strongly increased exciton decay time compared to a negative detuning 
(orange). This asymmetry has already been discussed in Ref.~\onlinecite{kaer:12} and is due to the
fact that at low temperatures phonon emission process dominate absorption processes and in fact at a
higher temperature (right panel) this asymmetry disappears. Here we have chosen to repeat these
kind of calculations to show that temperatures as high as $T=100$~K can easily be accessed using
the path-integral approach which is hard to reach with other methods.
Previously, the present model of a QD-cavity system has also been treated by
a hybrid approach\cite{vagov:14} which requires the knowledge of the phonon-induced
Rabi-frequency renormalization and the strength of the phonon-induced damping in the absence of 
cavity losses. These input parameters can be gained
by analyzing the damped oscillations of the excited state occupation in a lossless cavity occurring 
for constant resonant coupling as it can be calculated using the path-integral formalism in its
previous formulation\cite{vagov:11,glaessl:12b}. In the hybrid approach this input is then combined
with an \emph{a posteriori} phenomenological treatment of the cavity losses.
Thus, even though this way the damping and the frequency renormalization are 
calculated in a non-perturbative way the hybrid approach does not yet treat the cavity losses and the
phonon coupling on equal footing as the combined Lindblad and path-integral approach does.
Moreover, the derivation of the hybrid approach in Ref.~\onlinecite{vagov:14} made explicit use of
special properties that are only fulfilled for constant resonant excitation.
Therefore the present work also drastically extends the parameter range accessible with the
path-integral approach if one needs to include the effects of cavity losses, which in many cases
play a crucial role for the system dynamics.
Here, we have compared the results obtained from the hybrid approach for resonant excitation,
shown as black dots in Fig.~\ref{fig4}, with the path-integral method for both temperatures.
For low temperatures the results are identical, while at higher temperature small deviations
become visible. Similar to our previous comparison with the weak-coupling theory also here the
path-integral method serves as an excellent benchmark to explore the limits of the applicability
of less rigorous methods.


\section{Conclusions}\label{conclusions}  

We have presented an extension to the numerical path-integral formalism previously used to
calculate the reduced density matrix of a time-dependent few level system coupled to a set of
harmonic oscillator modes that allows a natural and systematic inclusion of non-Hamiltonian
dynamics within the path integral framework. Our combined method shows a way how to go beyond the 
representation of the path-integral formalism that relies on using the time-evolution operator and 
thus can be used to add arbitrary linear operations acting on the reduced density matrix to the 
equations of motion. 
We applied this new method to an optically coupled semiconductor quantum dot (QD) where the coupling
to longitudinal acoustic phonons is treated non-perturbatively for different optical excitation 
conditions and different environmental effects that are described in the Markov limit by 
corresponding phenomenological rates.
The combined Lindblad and path-integral method turns out to be a highly valuable tool for the 
treatment of optically coupled QDs with strong phonon interaction and similar quantum dissipative  
systems for a number of reasons. First of all, it allows to treat the deformation potential 
coupling to longitudinal acoustic phonons, which has been identified as the major decoherence 
mechanism in strongly confined QDs, in a numerically exact way that includes multi-phonon processes 
and all non-Markovian effects. This makes it possible to explore the regimes of arbitrarily strong 
QD-phonon coupling and both low and high temperatures in a non-perturbative way. Here, it is worth 
noting that within the path integral approach higher temperatures are actually easier to deal with 
as the memory time becomes shorter and thus less memory steps are needed while the scope of some 
approximate methods that explicitly truncate the phonon subspace is restricted to lower
temperatures. Besides the phonon coupling also the optical excitation can be chosen arbitrarily
as the formalism is able to deal with both weak and strong driving, constant and pulsed excitation
and also rapid changes of the excitation parameters that can prevent an adiabatic evolution of
the coupled light-matter system. This also includes situations with chirped excitation or with
multiple overlapping pulses of different frequency where the introduction of a suitable basis of 
photon-dressed states as needed by some schemes becomes non-obvious as the rotating frames naturally
associated with each pulse differ. Moreover, the method can be used to calculate all
elements of the reduced density matrix in the original frame of reference which not only gives
access to the QD occupations, but also to the coherences of the reduced system.
The inclusion of processes described by rate equations within the path-integral formalism 
that is made possible by the present work allows taking into account other loss channels that are 
relevant in typical experimental situations involving QDs. For example the presented method has
already been successfully used to include electron tunneling effects in photocurrent measurements
using off-resonant two-pulse and two-color excitation\cite{liu:16, brash:16}.

In our first application we discussed the combined effects of the radiative decay and the phonon
scattering on the driven stationary non-equilibrium state of a two-level QD for both resonant and
off-resonant excitation showing that there is a non-monotonic dependence of the stationary exciton
occupation on the driving strength. This is not expected in both limiting cases where
either the dot phonon interaction or the radiative decay provide the only environment coupling.
So far no unbiased approach was formulated for studying such
combination effects. We then used the path-integral results to explore the range of validity of a
weak-coupling method that treats the phonon coupling perturbatively and found for strong
driving in the regime of detuned excitation at high temperatures significant deviations between 
the two methods while otherwise the weak-coupling theory works well over wide parameter ranges.
Finally, we applied our new formalism to the case of a QD coupled to a single cavity mode and 
analyzed the exciton lifetime that is limited due to photon losses of the cavity for different
detunings of the optical mode from the
QD resonance at low and very high temperatures. The combined Lindblad and path-integral method has
also been used as a benchmark to test a previously developed hybrid approach that was limited to
resonant excitation.

\section{ACKNOWLEDGMENTS}

A.M.B. and V.M.A. gratefully acknowledge the financial support from Deutsche Forschungsgemeinschaft
via the Project No. AX 17/7-1.

\end{document}